\documentclass[12pt]{article}

\usepackage[margin=1in,dvips,a4paper,nohead]{geometry}
\usepackage[margin=1em, font=small, justification=centerlast, tableposition=top]{caption}
\usepackage{float}

\usepackage{amsfonts,amssymb,amsmath}
\allowdisplaybreaks
\usepackage[noBBpl,slantedGreek]{mathpazo}

\usepackage{graphicx}

\usepackage{pstricks}

\def \bbZ {\mathbb Z}
\def \calP {\mathcal P}
\def \calQ {\mathcal Q}
\def \rme {\mathrm e}
\def \rmi {\mathrm i}

\def \phmo {\hbox to 0pt{\hss \phantom{\scriptsize -1}}}

\begin{document}

\title{Three-coloring statistical model with\\ domain wall boundary conditions.\\ II. Trigonometric limit}
\author{A.~V.~Razumov, Yu.~G.~Stroganov\\
\small \it Institute for High Energy Physics\\[-.5em]
\small \it 142281 Protvino, Moscow region, Russia}
\date{}

\maketitle

\begin{abstract}
A nontrivial trigonometric limit of the three-coloring statistical model with the domain wall boundary conditions is considered. In this limit the functional equations, constructed in the previous paper, are solved and a new determinant representation for the partial partition functions is found.
\end{abstract}

\section{Introduction}

In this paper we continue investigation of the statistical three-coloring model  with the domain wall boundary conditions started in our paper \cite{RazStr08}. The model is directly related to the Baxter's three-coloring model \cite{Bax70, Bax82}. First, we recall the necessary definitions and facts.

The states of the model are various colorings of an $(n + 1) \times (n + 1)$ grid of faces with three colors, such that any two adjacent faces have different colors. It is convenient to label the colors by the elements of the ring $\bbZ_3$.\footnote{We denote the elements of $\bbZ_3$ by $\overline 0$, $\overline 1$ and $\overline 2$. When an element of $\bbZ_3$ arises in a context where an integer should be, it is treated as the corresponding integer $0$, $1$ or $2$. Vice verse, when an integer $i$ arises in a context where an element of $\bbZ_3$ should be, it is treated as the element of $\bbZ_3$ corresponding to the reminder after division of $i$ by 3.} Here, the colors of adjacent faces are different if and only if the corresponding labels differ by $+\overline 1$ or $-\overline 1$. The domain wall boundary conditions means that if one starts with any boundary face and walks anticlockwise along the boundary then the color changes by $+ \overline 1$ from face to face for the vertical boundaries, and by $- \overline 1$ for the horizontal boundaries.  An example is given in Figure \ref{DomainWallThreeColoring}.
\begin{figure}[htb]
\centering
\begin{pspicture}(0,-.8)(5.8,5.2)
\multips(0,0)(0,1){6}{
  \psline(0,0)(5,0)}
\multips(0,0)(1,0){6}{
  \psline(0,0)(0,5)  }
\multips(0,0)(0,1){6}{
  \multips(0,0)(1,0){6}{
    \pscircle[fillstyle=solid](0,0){.08}}}
\rput(.5,4.5){$\scriptstyle \overline 2$}
\rput(1.5,4.5){$\scriptstyle \overline 0$}
\rput(2.5,4.5){$\scriptstyle \overline 1$}
\rput(3.5,4.5){$\scriptstyle \overline 2$}
\rput(4.5,4.5){$\scriptstyle \overline 0$}
\rput(4.5,3.5){$\scriptstyle \overline 2$}
\rput(4.5,2.5){$\scriptstyle \overline 1$}
\rput(4.5,1.5){$\scriptstyle \overline 0$}
\rput(4.5,0.5){$\scriptstyle \overline 2$}
\rput(.5,3.5){$\scriptstyle \overline 0$}
\rput(.5,2.5){$\scriptstyle \overline 1$}
\rput(.5,1.5){$\scriptstyle \overline 2$}
\rput(.5,.5){$\scriptstyle \overline 0$}
\rput(1.5,.5){$\scriptstyle \overline 2$}
\rput(2.5,.5){$\scriptstyle \overline 1$}
\rput(3.5,.5){$\scriptstyle \overline 0$}
\rput(1,-.4){$\psi_1$}
\rput(2,-.4){$\psi_2$}
\rput(3,-.4){$\psi_3$}
\rput(4,-.4){$\psi_4$}
\rput(5.5,4){$\chi_1$}
\rput(5.5,3){$\chi_2$}
\rput(5.5,2){$\chi_3$}
\rput(5.5,1){$\chi_4$}
\end{pspicture}
\caption{}
\label{DomainWallThreeColoring}
\end{figure}

The Boltzmann weight of a state is the product of the Boltzmann weights of the internal vertices of the lattice, and the sum of the weights of all possible states is the partition function of the model. We assume that the Boltzmann weight of a vertex  is determined by the colors of the four adjacent faces. To have an integrable system, we assume that they also depend on  a spectral parameter so that the star-triangle relation (Yang--Baxter equations) is satisfied. It is not difficult to get convinced that there are six types of possible vertex color configurations  given in Figure~\ref{ThreeColorings},
\begin{figure}[htb]
\centering
\begin{tabular}{cccccc}
\begin{pspicture}(-.1,-.1)(2.1,2.1)
\psline (0,0)(2,0)
\psline (0,1)(2,1)
\psline (0,2)(2,2)
\psline (0,0)(0,2)
\psline (1,0)(1,2)
\psline (2,0)(2,2)
\multips(0,0)(0,1){3}{
  \multips(0,0)(1,0){3}{
    \pscircle[fillstyle=solid](0,0){.08}}}
\rput(.5,1.5){$\scriptstyle r - 1$}
\rput(1.5,1.5){$\scriptstyle r$}
\rput(1.5,.5){$\scriptstyle r + 1$}
\rput(.5,.5){$\scriptstyle r$}
\end{pspicture} &
\begin{pspicture}(-.1,-.1)(2.1,2.1)
\psline (0,0)(2,0)
\psline (0,1)(2,1)
\psline (0,2)(2,2)
\psline (0,0)(0,2)
\psline (1,0)(1,2)
\psline (2,0)(2,2)
\multips(0,0)(0,1){3}{
  \multips(0,0)(1,0){3}{
    \pscircle[fillstyle=solid](0,0){.08}}}
\rput(.5,1.5){$\scriptstyle r + 1$}
\rput(1.5,1.5){$\scriptstyle r$}
\rput(1.5,.5){$\scriptstyle r - 1$}
\rput(.5,.5){$\scriptstyle r$}
\end{pspicture} &
\begin{pspicture}(-.1,-.1)(2.1,2.1)
\psline (0,0)(2,0)
\psline (0,1)(2,1)
\psline (0,2)(2,2)
\psline (0,0)(0,2)
\psline (1,0)(1,2)
\psline (2,0)(2,2)
\multips(0,0)(0,1){3}{
  \multips(0,0)(1,0){3}{
    \pscircle[fillstyle=solid](0,0){.08}}}
\rput(.5,1.5){$\scriptstyle r$}
\rput(1.5,1.5){$\scriptstyle r - 1$}
\rput(1.5,.5){$\scriptstyle r$}
\rput(.5,.5){$\scriptstyle r + 1$}
\end{pspicture} &
\begin{pspicture}(-.1,-.1)(2.1,2.1)
\psline (0,0)(2,0)
\psline (0,1)(2,1)
\psline (0,2)(2,2)
\psline (0,0)(0,2)
\psline (1,0)(1,2)
\psline (2,0)(2,2)
\multips(0,0)(0,1){3}{
  \multips(0,0)(1,0){3}{
    \pscircle[fillstyle=solid](0,0){.08}}}
\rput(.5,1.5){$\scriptstyle r$}
\rput(1.5,1.5){$\scriptstyle r + 1$}
\rput(1.5,.5){$\scriptstyle r$}
\rput(.5,.5){$\scriptstyle r - 1$}
\end{pspicture} &
\begin{pspicture}(-.1,-.1)(2.1,2.1)
\psline (0,0)(2,0)
\psline (0,1)(2,1)
\psline (0,2)(2,2)
\psline (0,0)(0,2)
\psline (1,0)(1,2)
\psline (2,0)(2,2)
\multips(0,0)(0,1){3}{
  \multips(0,0)(1,0){3}{
    \pscircle[fillstyle=solid](0,0){.08}}}
\rput(.5,1.5){$\scriptstyle r$}
\rput(1.5,1.5){$\scriptstyle r + 1$}
\rput(1.5,.5){$\scriptstyle r$}
\rput(.5,.5){$\scriptstyle r + 1$}
\end{pspicture} &
\begin{pspicture}(-.1,-.1)(2.1,2.1)
\psline (0,0)(2,0)
\psline (0,1)(2,1)
\psline (0,2)(2,2)
\psline (0,0)(0,2)
\psline (1,0)(1,2)
\psline (2,0)(2,2)
\multips(0,0)(0,1){3}{
  \multips(0,0)(1,0){3}{
    \pscircle[fillstyle=solid](0,0){.08}}}
\rput(.5,1.5){$\scriptstyle r$}
\rput(1.5,1.5){$\scriptstyle r - 1$}
\rput(1.5,.5){$\scriptstyle r$}
\rput(.5,.5){$\scriptstyle r - 1$}
\end{pspicture} \\[1em]
$\alpha_r(\varphi)$ & $\alpha'_r(\varphi)$ & $\beta_r(\varphi)$ & $\beta'_r(\varphi)$ & $\gamma_r(\varphi)$ & $\gamma'_r(\varphi)$
\end{tabular}
\caption{}
\label{ThreeColorings}
\end{figure}
where $r$ is one of the colors $\overline 0$, $\overline 1$, and $\overline 2$.  We will use the weights found by Stroganov \cite{Str82} and having the form\footnote{Actually, the weights (\ref{1})--(\ref{4}) are connected with the weights found by Stroganov by a gauge transformation. In the paper \cite{RazStr08} we mark these weights and the corresponding partition function by tilde. To simplify notation, we omit the corresponding tildes in the present paper.}
\begin{align}
&\alpha_r(\varphi) = \alpha'_r(\varphi) = \frac{\theta_1(\pi / 3 - \varphi)}{\theta_1(2 \pi / 3)}, \label{1} \\
& \beta_r(\varphi) = \beta'_r(\varphi) = \zeta_r^{1/2} \,  \frac{\theta_1(\pi / 3 + \varphi)}{\theta_1(2 \pi / 3)}, \\
&\gamma_r(\varphi) = \frac{\theta_4(\lambda + 2 \pi (r + 1 / 2) / 3 + \varphi)}{\theta_4(\lambda + 2 \pi r  / 3)}, \\
& \gamma{}'_r(\varphi) = \frac{\theta_4(\lambda + 2 \pi (r - 1 / 2) / 3 - \varphi)}{\theta_4(\lambda + 2 \pi r / 3)}. \label{4}
\end{align}
Here $\theta_1$ and $\theta_4$ are standard elliptic $\theta$-functions of nome $p = \rme^{\rmi \pi \tau}$~\cite{WhiWat27}, $\varphi$ is the spectral parameter, $\lambda$ is a fixed parameter, and
\begin{equation*}
\zeta_r = \frac{\theta_4(\lambda + 2 \pi (r - 1) / 3) \theta_4(\lambda +  2 \pi (r + 1) / 3)}{\theta_4^2(\lambda + 2 \pi r / 3)}.
\end{equation*}

We will consider the inhomogeneous case when the internal horizontal lines are labeled by the variables $\chi_i$, $i = 1, \ldots, n$, and the internal vertical lines are labeled by the variables  $\psi_i$, $i = 1, \ldots, n$, see for example Figure~\ref{DomainWallThreeColoring}. With the vertex at the intersection of the line labeled by $\chi_i$ and the line labeled by $\psi_j$ we associate the spectral parameter $\chi_i - \psi_j$. The total partition function $Z_n(\{\chi\}; \{\psi\})$ is the sum of partial partition functions:
\[
Z_n(\{\chi\}; \{\psi\}) = \sum_{r \in \bbZ_3} Z^r_n(\{\chi\}; \{\psi\}),
\]
where  $r$ is the color of the left topmost vertex of the lattice.

Now we introduce the functions
\begin{multline*}
F^r_n(\{\chi\};  \{\psi\}) = \frac{1}{\theta_4(\lambda + 2 \pi (r + n) / 3)}  \\*
\times \prod_{\substack{i,j = 1 \\ i < j}}^n \theta_1 (\chi_i - \chi_j) \prod_{i,j = 1}^n \theta_1(\chi_i - \psi_j)  \prod_{\substack{i,j = 1 \\ i < j}}^n \theta_1 (\psi_i - \psi_j) \, Z^r_n(\{\chi\}; \{\psi\}).
\end{multline*}
It was proved in the paper \cite{RazStr08} that the functions $F^r_n(\{\chi\};  \{\psi\})$ satisfy the functional equations
\begin{equation}
\sum_{s \in \bbZ_3} F_n^{r + s}(\chi_1, \ldots, \chi_k + 2 \pi s / 3, \ldots, \chi_n; \{\psi\}) = 0, \label{f:1}
\end{equation}
and the functional equations
\begin{equation}
\sum_{s \in \bbZ_3} F_n^{r + s}(\{\chi\}; \psi_1, \ldots, \psi_k - 2 \pi s / 3,  \ldots, \psi_n) = 0. \label{f:2}
\end{equation}
The above equations are similar to the equations obtained by Stroganov \cite{Str06} for the six-vertex model with the domain wall boundary conditions for the special value of the crossing parameter $\eta = 2 \pi / 3$. It appeared \cite{Str06}  that these equations are very useful for solving enumeration problems related to alternating sign matrices, see also the papers \cite{RazStr04, RazStr06a, RazStr06b}.

In the present paper we solve the equations (\ref{f:1}) and (\ref{f:2}) in the trigonometric limit and find a determinant representation for the partial partition functions. Note that the three-coloring statistical model is a partial case of the eight-vertex-solid-on-solid-model, introduced by Baxter \cite{Bax73}. A determinant representation of the partition function of this model for the case of the domain wall boundary conditions was obtained by Rosengren~\cite{Ros08}. In principle, our representation can be obtained from the Rosengren's representation, using Okada's results on determinants  \cite{Oka98, Oka06}. However, we use the functional equations and hope that the present consideration will be useful for consideration of the general elliptic case.

\section{Trigonometric limit}

When $\operatorname{Im} \tau \to \infty$ ($p \to 0$) the weights (\ref{1})--(\ref{4}) take the form
\begin{gather*}
\alpha_r(\varphi) = \alpha'_r(\varphi) = \frac{\sin(\pi / 3 - \varphi)}{\sin(2 \pi / 3)}, \qquad
\beta_r(\varphi) = \beta'_r(\varphi) =  \frac{\sin(\pi / 3 + \varphi)}{\sin(2 \pi / 3)}, \\[.5em]
\gamma_r(\varphi) = \gamma'_r(\varphi) = 1.
\end{gather*}
It is used here that at $p \to 0$ we have
\[
\theta_1(\varphi | p) = 2 p^{1/4} \sin \varphi + O(p^{9/4}), \qquad \theta_4(\varphi | p) = 1 + O(p).
\]
Thus, we come to the weights of the six-vertex model for a special value of the crossing parameter $\eta = 2 \pi / 3$ and see that each partial partition function of the three-coloring statistical model coincide with the partition function of the six-vertex model for this value of the crossing parameter.

There is also a nontrivial trigonometric limit. To obtain it, we make first the substitution $\lambda \to \lambda + \pi \tau / 2$ and use the relation
\[
\theta_4(\varphi | p) = \rmi p^{1/4} \rme^{- \rmi \varphi} \theta_1(\varphi - \pi \tau / 2 | p).
\]
The weights (\ref{1})--(\ref{4}) take now the form
\begin{align*}
&\alpha_r(\varphi) = \alpha'_r(\varphi) = \frac{\theta_1(\pi / 3 - \varphi)}{\theta_1(2 \pi / 3)}, \\
& \beta_r(\varphi) = \beta'_r(\varphi) = \zeta_r^{1/2} \, \frac{\theta_1(\pi / 3 + \varphi)}{\theta_1(2 \pi / 3)}, \\
&\gamma_r(\varphi) = \rme^{-\rmi (\pi / 3 + \varphi)} \, \frac{\theta_1(\lambda + 2 \pi (r + 1 / 2) / 3 + \varphi)}{\theta_1(\lambda + 2 \pi r  / 3)}, \\
& \gamma{}'_r(\varphi) = \rme^{\rmi (\pi / 3 + \varphi)} \, \frac{\theta_1(\lambda + 2 \pi (r - 1 / 2) / 3 - \varphi)}{\theta_1(\lambda + 2 \pi r / 3)},
\end{align*}
where
\begin{equation*}
\zeta_r = \frac{\theta_1(\lambda + 2 \pi (r - 1) / 3) \theta_1(\lambda +  2 \pi (r + 1) / 3)}{\theta_1^2(\lambda + 2 \pi r / 3)}.
\end{equation*}
We prefer to work with polynomials rather than with trigonometric functions, therefore we denote
\[
w = \rme^{\rmi \varphi}, \qquad a = \rme^{\rmi \pi / 3}, \qquad b = \rme^{\rmi \lambda},
\]
and
\[
\sigma(w) = w - w^{-1}.
\]
Then, for $p \to 0$ we obtain
\begin{align}
& \alpha_r(w) = \alpha'_r(w) = \frac{\sigma(a \, w^{-1})}{\sigma(a^2)}, && \beta_r(u) = \beta'_r(u) = \zeta_r^{1/2} \, \frac{\sigma(a \, w)}{\sigma(a^2)}, \label{5} \\
& \gamma_r(w) = a^{-1} \, w^{-1} \, \frac{\sigma(a^{2r + 1} b \, w)}{\sigma(a^{2r} b)}, && \gamma'_r(u) = a \, w \, \frac{\sigma(a^{2r - 1} b \, w^{-1})}{\sigma(a^{2r} b)}, \label{6}
\end{align}
where
\[
\zeta_r = \frac{\sigma(a^{2(r - 1)} b) \, \sigma(a^{2(r + 1)} b)}{\sigma^2(a^{2r} b)}.
\]

Before going to the trigonometric limit for the functions $F^r_n(\{x\};  \{y\})$, we divide them by $(- \rmi p^{1/4})^{n(2n - 1)}$ retaining the notation. It is not difficult to see that at $p \to 0$ we have
\[
F^r_n(\{x\};  \{y\}) = \frac{a^{2(r + n)}b}{\sigma(a^{2(r + n)}b)} \prod_{\substack{i,j = 1 \\ i < j}}^n \sigma(x_i^{\phmo} x_j^{-1}) \prod_{i,j = 1}^n \sigma(x_i^{\phmo} y_j^{-1})  \prod_{\substack{i,j = 1 \\ i < j}}^n \sigma(y_i^{\phmo} y_j^{-1}) \, Z^r_n(\{x\}; \{y\}),
\]
where
\[
x_i = \rme^{\rmi \chi_i}, \qquad y_i = \rme^{\rmi \psi_i}.
\]
Now the functional equations (\ref{f:1}) and (\ref{f:2}) can be rewritten as\footnote{In accordance with our convention, we assume that $a^{\pm 2 \overline m} = a^{\pm 2 m}$.}
\begin{gather}
\sum_{s \in \bbZ_3} F_n^{r + s}(x_1, \ldots, a^{2s} x_k, \ldots, x_n; \{y\}) = 0, \label{f:3} \\
\sum_{s \in \bbZ_3}^2 F_n^{r + s}(\{x\}; y_1, \ldots, a^{-2s} y_k,  \ldots, y_n) = 0. \label{f:4}
\end{gather}
In the next section we solve these equations.

\section{Solving functional equations}

For a fixed $n$, each of the functional equations (\ref{f:3}) and (\ref{f:4}) contains all three functions $F_n^s(\{x\}, \{y\})$. It is convenient to construct functional equations each depending on one function. To this end we make the discrete Fourier transformation,
\[
W_n^r(\{x\}, \{y\}) = \sum_{s \in \bbZ_3} a^{-2rs} F_n^s(\{x\}, \{y\}),
\]
whose inverse is
\[
F_n^r(\{x\}, \{y\}) = \frac{1}{3} \sum_{s \in \bbZ_3} a^{2 rs} W_n^s(\{x\}, \{y\}).
\]
The functional equations (\ref{f:3}) and (\ref{f:4}) in terms of the functions $W_n^s(\{x\}, \{y\})$ take the form
\begin{gather}
\sum_{s \in \bbZ_3} a^{2 rs} W_n^r(x_1, \ldots, a^{2s} x_k, \ldots, x_n; \{y\}) = 0, \label{7} \\
\sum_{s \in \bbZ_3} a^{2 rs} W_n^r(\{x\}; y_1, \ldots, a^{-2s} y_k,  \ldots, y_n) = 0. \label{8}
\end{gather}

Having in mind the correspondence between the states of the three-coloring model and the six-vertex model, found by Lenard \cite{Len61}, we conclude that for any state of the three-coloring model in each column of vertices there is at least one vertex of type $\gamma$, and the number of vertex of type $\gamma$ is greater by one than the number of states of type~$\gamma'$. Using the explicit form of the weights in the trigonometric limit (\ref{5})--(\ref{6}), we see that
\begin{align}
W_n^r(x_1, \ldots, - x_k, \ldots, x_n; \{y\}) &= (-1)^n W_n^r(x_1, \ldots, x_k, \ldots, x_n; \{y\}), \label{9} \\[.5em]
W_n^r(\{x\}; y_1, \ldots, - y_k, \ldots, y_n) &= (-1)^n W_n^r(\{x\}; y_1, \ldots, y_k, \ldots, y_n), \label{10}
\end{align}
and that we can write
\begin{align}
W_n^r(x_1, \ldots, x_k, \ldots, x_n; \{y\}) &= \sum_{\ell = 1}^{3n} \alpha^r_{n,k,l}(x_1, \ldots, \widehat{x_k}, \ldots, x_n; \{y\}) x_k^{3n - 2 \ell}, \label{11} \\
W_n^r( \{x\}, y_1, \ldots, y_k, \ldots, y_n;) &= \sum_{\ell = 0}^{3n-1} \beta^r_{n,k,l}(\{x\}, y_1, \ldots, \widehat{y_k}, \ldots, y_n; \{y\}) y_k^{3n - 2 \ell}, \label{12}
\end{align}
where the hat denotes omission of the corresponding argument.

The form of the functional equations (\ref{7}) and (\ref{8}) suggests to introduce the functions
\[
V_n^r(\{x\}, \{y\}) = \prod_{i=1}^n \left( x_i^r y_i^{-r} \right) W_n^r(\{x\}, \{y\}),
\]
which satisfy simpler functional equations
\[
\sum_{s \in \bbZ_3} V_n^r(x_1, \ldots, a^{2s} x_k, \ldots, x_n; \{y\}) = 0, \qquad
\sum_{s \in \bbZ_3} V_n^r(\{x\}; y_1, \ldots, a^{2s} y_k,  \ldots, y_n) = 0.
\]
Now instead of the relations (\ref{9}) and  (\ref{10}) we have the relations
\begin{align*}
V_n^r(x_1, \ldots, - x_k, \ldots, x_n; \{y\}) &= (-1)^{3n+r} V_n^r(x_1, \ldots, x_k, \ldots, x_n; \{y\}), \\[.5em]
V_n^r(\{x\}; y_1, \ldots, - y_k, \ldots, y_n) &= (-1)^{3n-r} V_n^r(\{x\}; y_1, \ldots, y_k, \ldots, y_n),
\end{align*}
and instead of the representations  (\ref{11}) and  (\ref{12}) we have the representations
\begin{align}
V_n^r(x_1, \ldots, x_k, \ldots, x_n; \{y\}) &= \sum_{\ell = 1}^{3n} \gamma^r_{n,k,l}(x_1, \ldots, \widehat{x_k}, \ldots, x_n; \{y\}) x_k^{3n - 2 \ell + r}, \label{13} \\
V_n^r( \{x\}, y_1, \ldots, y_k, \ldots, y_n;) &= \sum_{\ell = 0}^{3n-1} \delta^r_{n,k,l}(\{x\}, y_1, \ldots, \widehat{y_k}, \ldots, y_n; \{y\}) y_k^{3n - 2 \ell - r}. \label{14}
\end{align}

Now we consider the properties of the functions $V_n^r(\{x\}, \{y\})$ for different $r$ separately. We start with $r = \overline 0$. In this case the functional equations (\ref{13}) and (\ref{14}) give
\begin{align*}
V_n^{\overline 0}(x_1, \ldots, x_k, \ldots, x_n; \{y\}) = \sum_{\substack{ \ell = 1 \\ 3 \, \nmid \, \ell}}^{3n - 1} \gamma^{\overline 0}_{n,k,l}(x_1, \ldots, \hat x_k, \ldots, x_n; \{y\}) x_k^{3n - 2 \ell}, \\
V_n^{\overline 0}(\{y\}; y_1, \ldots, y_k, \ldots, y_n; ) = \sum_{\substack{ \ell = 1 \\ 3 \, \nmid \, \ell}}^{3n - 1} \delta^{\overline 0}_{n,k,l}(\{x\}; y_1, \ldots, \hat y_k, \ldots, y_n) y_k^{3n - 2 \ell}.
\end{align*}
We introduce the notation
\[
u_{2i - 1} = x_i, \qquad u_{2i} = y_{i}, \qquad i = 1, \ldots, n,
\]
and summarize the properties of the functions $V_n^{\overline 0}(\{u\})$.
\begin{itemize}
\item For every $\mu = 1, \ldots, 2n$ the function $V_n^{\overline 0}(\{u\})$ satisfies the functional equations
\begin{multline*}
V_n^{\overline 0}(u_1, \ldots, u_\mu, \ldots, u_{2n}) + V_n^{\overline 0}(u_1, \ldots, a^2 u_\mu, \ldots, u_{2n}) \\+ V_n^{\overline 0}(u_1, \ldots, a^4 u_\mu, \ldots, u_{2n}) = 0.
\end{multline*}
\item For every $\mu = 1, \ldots, 2n$, the function $u_\mu^{3n-2} V_n^{\overline 0}(\{u\})$ is a polynomial in $u_\mu^2$ of degree not greater then $3n - 2$.
\item The function $V_n^{\overline 0}(\{u\})$ vanishes if $u_\mu^2 = u_\nu^2$ for some $\mu \ne \nu$.
\end{itemize}
It can be shown \cite{Str06, RazStr04} that a function, satisfying the three above properties, is proportional to the determinant of the matrix
\[
P_n(\{u\}) = \left(
\begin{array}{ccccc}
u_1^{3n - 2} & u_2^{3n - 2} & u_3^{3n - 2} & \cdots & u_{2n}^{3n - 2} \\
u_1^{3n - 4} & u_2^{3n - 4} & u_3^{3n - 4} & \cdots & u_{2n}^{3n - 4} \\
u_1^{3n - 8} & u_2^{3n - 8} & u_3^{3n - 8} & \cdots & u_{2n}^{3n - 8} \\
\vdots & \vdots & \vdots & \ddots & \vdots \\
u_1^{- 3n + 2} & u_2^{- 3n + 2} & u_3^{- 3n + 2} & \cdots & u_{2n}^{- 3n + 2}
\end{array}
\right).
\]

In a similar way we can summarize the properties of the functions $V_n^{\overline 1}(\{u\})$ as follows.
\begin{itemize}
\item For every $\mu = 1, \ldots, 2n$ the function $V_n^{\overline 1}(\{u\})$ satisfies the functional equations
\begin{multline*}
V_n^{\overline 1}(u_1, \ldots, u_\mu, \ldots, u_{2n}) + V_n^{\overline 1}(u_1, \ldots, a^2 u_\mu, \ldots, u_{2n}) \\+ V_n^{\overline 1}(u_1, \ldots, a^4 u_\mu, \ldots, u_{2n}) = 0.
\end{multline*}
\item For every $\mu = 1, \ldots, 2n$, the function $u_\mu^{3n-1} V_n^{\overline 1}(\{u\})$ is a polynomial in $u_\mu^2$ of degree not greater then $3n - 1$.
\item The function $V_n^{\overline 1}(\{u\})$ vanishes if $u_\mu^2 = u_\nu^2$ for some $\mu \ne \nu$.
\end{itemize}
It can be shown \cite{Str04} that a function, satisfying these three properties, is proportional to the determinant of the matrix
\[
Q_n(\{u\}) = \left(
\begin{array}{ccccc}
u_1^{3n - 1} & u_2^{3n - 1} & u_3^{3n - 1} & \cdots & u_{2n}^{3n - 1} \\
u_1^{3n - 5} & u_2^{3n - 5} & u_3^{3n - 4} & \cdots & u_{2n}^{3n - 5} \\
u_1^{3n - 7} & u_2^{3n - 7} & u_3^{3n - 7} & \cdots & u_{2n}^{3n - 7} \\
\vdots & \vdots & \vdots & \ddots & \vdots \\
u_1^{- 3n + 1} & u_2^{- 3n + 1} & u_3^{- 3n + 1} & \cdots & u_{2n}^{- 3n + 1}
\end{array}
\right).
\]

Finally, the functions $V_n^{\overline 2}(\{u\})$ satisfy the same properties as the functions $V_n^{\overline 0}(\{u\})$. Hence, they are also proportional to the determinant of the matrices $P_n(\{u\})$.

We do not give here the expressions for the proportionality coefficients of the functions $V_n^r(\{u\})$ and the determinants of the matrices $P(\{u\})$ and $Q(\{u\})$, but proceed directly to the consideration of the partial partition functions.

\section{Partition functions}

Now we find an explicit expression for the partial partition functions. Note first that in the general elliptic case the partial partition functions of the model satisfy some recursion relations \cite{RazStr08, Ros08}. In particular, we have
\begin{multline*}
\left. Z^r_n(\chi_1, \ldots, \chi_{n-1}, \chi_n; \psi_1, \ldots, \psi_{n-1}, \psi_n) \right|_{\chi_n = \psi_n + \pi / 3} =  \theta_1^{2 - 2n}(2 \pi / 3) \\
\times \frac{\theta_4(\lambda + 2 \pi (r + n) / 3)} {\theta_4(\lambda + 2 \pi (r + n - 1) / 3)}  \prod_{i = 1}^{n-1} \theta_1(\chi_i - \psi_n  + \pi / 3) \prod_{i = 1}^{n-1} \theta_1(\psi_n - \psi_i + 2 \pi / 3) \\*
\times \widetilde Z^r_{n-1}(\chi_1, \ldots, \chi_{n-1}; \psi_1, \ldots, \psi_{n-1}).
\end{multline*}
In the trigonometric limit these relations take the form
\begin{multline*}
Z^r_n(u_1, \ldots, u_{2n-1}, u_{2n}) |_{u_{2n} = a^{-1} u_{2n - 1}} \\
= a^{-2} \sigma^{2-2n}(a^2) \frac{\sigma(a^{2(r + n)} b)}{\sigma(a^{2(r + n - 1)} b)} \prod_{\mu = 1}^{2n - 2} \sigma(a^{-1} u_\mu^{\phmo} u_{2n}^{-1}) \, Z^r_{n-1}(u_1, \ldots, u_{2n - 2}).
\end{multline*}
It is convenient to introduce the functions
\[
Z^{\prime r}_n(\{u\}) = \frac{a^{2(n-1)} \sigma^{n(n-1)}(a^2)}{\sigma(a^{2(n + r)} b)} \; Z_{\mu, n}(\{u\}),
\]
which satisfy simpler recursion relations
\begin{multline}
Z'_{\mu, n}(u_1, \ldots, u_{2n-2}, u_{2n-1}, u_{2n}) |_{u_{2n} = a^{-1} u_{2n - 1}}
\\= \prod_{\mu=1}^{2 n-2} \sigma(a^{-1} u_\mu^{\phmo} u_{2n - 1}^{-1}) Z'_{\mu, n-1}(u_1, \ldots, u_{2n-2}). \label{17}
\end{multline}

Having in mind that the partial partition functions can be expressed via the functions $V_n^r(\{u\})$, we see that the functions $Z_n^{\prime r}(\{u\})$ can be represented as
\begin{equation}
Z_n^{\prime r}(\{u\}) = A_n^r \calP_n(\{u\}) + B_n^r \prod_{i=1}^n (u_{2 i-1}^{-1} u_{2i}^{\phmo}) \calQ_n(\{u\}) + C_n^r \prod_{i=1}^n (u_{2 i-1}^{-2} u^2_{2i}) \calP_n(\{u\}), \label{18}
\end{equation}
where
\begin{gather*}
\calP_n(\{u\}) = \frac{1}{\prod_{1 \le \mu < \nu \le 2n} \sigma(u_\mu^{\phmo} u_\nu^{-1})} \det P_n(\{u\}), \\
\calQ_n(\{u\}) = \frac{1}{\prod_{1 \le \mu < \nu \le 2n} \sigma(u_\mu^{\phmo} u_\nu^{-1})} \det Q_n(\{u\}),
\end{gather*}
and $A_n^r$, $B_n^r$, $C_n^r$ are some constants. Note that the functions $\calP_n(\{u\})$ and $\calQ_n(\{u\})$ are symmetric functions in the variables $u_1$, $\ldots$, $u_{2n}$. The functions $\calP_n(\{u\})$ are directly related to certain Schur functions,
\[
\calP_n(\{u\}) = (u_1 \cdots u_{2n})^{-n} s_{(n-1, \, n-1, \, n-2, \, n-2, \ldots, 1, \, 1, \, 0, \, 0)}(u_1^2, \ldots, u_{2n}^2),
\]
that was actually remarked by Okada~\cite{Oka06}. For the functions $\calQ_n(\{u\})$ we have
\[
\calQ_n(\{u\}) = (u_1 \cdots u_{2n})^{-n-1} s_{(n, \, n-1, \, n-1, \, n-2, \ldots, 2, \, 1, \, 1, \, 0)}(u_1^2, \ldots, u_{2n}^2).
\]

It can be shown that the functions $\det P_n(\{u\})$ and $\det Q_n(\{u\})$ satisfy the recursion relations
\begin{align}
\det P_n(u_1, \ldots, u_{2n - 1}, &u_{2n})|_{u_{2n} = a^{-1} u_{2n - 1}} \notag \\
&= (-1)^{n-1} \sigma(a) \prod_{\mu=1}^{2n - 2} \sigma(u_\mu^3 u_{2n-1}^{-3}) \; \det P_{n-1}(u_1, \ldots, u_{2n - 2}), \label{15} \\
\det Q_n(u_1, \ldots, u_{2n - 1}, &u_{2n})|_{u_{2n} = a^{-1} u_{2n - 1}} \notag \\*
&= (-1)^{n-1} \sigma(a) \prod_{\mu=1}^{2n - 2} \sigma(u_\mu^3 u_{2n-1}^{-3}) \; \det Q_{n-1}(u_1, \ldots, u_{2n - 2}). \label{16}
\end{align}
For the proof of the first relation we refer the reader to our paper \cite{RazStr04}. The second relations can be proved similarly. The recursion relations (\ref{15}) and
(\ref{16}) imply the recursion relation for the functions $\calP(\{u\})$ and $\calQ(\{u\})$,
\begin{align}
\calP_n(u_1, \ldots, u_{2n-2}, u_{2n-1}, &u_{2n}) |_{u_{2n} = a^{-1} u_{2n - 1}} \notag \\
&= (-1)^{n-1} \prod_{\mu=1}^{2 n-2} \sigma(a^{-1} u_\mu^{\phmo} u_{2n - 1}^{-1}) \calP_{n-1}(u_1, \ldots, u_{2n-2}), \label{19} \\
\calQ_n(u_1, \ldots, u_{2n-2}, u_{2n-1}, &u_{2n}) |_{u_{2n} = a^{-1} u_{2n - 1}} \notag \\
&= (-1)^{n-1} \prod_{\mu=1}^{2 n-2} \sigma(a^{-1} u_\mu^{\phmo} u_{2n - 1}^{-1}) \calQ_{n-1}(u_1, \ldots, u_{2n-2}). \label{20}
\end{align}
Using the representation (\ref{18}), the recursion relations (\ref{17}), (\ref{19}) and (\ref{20}), we obtain the following recursion relations for the coefficients
\begin{equation}
A^r_n = (-1)^{n - 1} A^r_{n-1}, \qquad B^r_n = (-1)^{n - 1} a B^r_{n-1}, \qquad C^r_n = (-1)^{n - 1} a^2 C^r_{n-1}. \label{21}
\end{equation}

Deriving these recursion relations we assumed that the three functions entering the right hand side of the equality (\ref{18}) are linearly independent. It is not difficult to see that it is not the case for $n = 1$. Hence, the recursion relations (\ref{21}) are actually valid for $n > 2$. After some quite lengthy calculations we find
\[
A_2^r = - \frac{a^{4r - 2} \, b^2}{\sigma(b^3)}, \qquad B_2^r = - \frac{1}{\sigma(b^3)}, \qquad C_2^r = - \frac{a^{2 - 4r} \, b^{-2}}{\sigma(b^3)}.
\]
Using these relations as initial conditions for the recursion relations (\ref{21}), we find
\begin{gather*}
A^r_n = (-1)^{n(n - 1)/2} \, \frac{a^{4r - 2} \, b^2}{\sigma(b^3)}, \qquad B^r_n = (-1)^{n(n - 1)/2} \, \frac{a^{n-2}}{\sigma(b^3)}, \\
C^r_n = (-1)^{n(n - 1)/2} \, \frac{a^{2n - 4r - 2} \, b^{-2}}{\sigma(b^3)}.
\end{gather*}
Note that these expressions work for all $n \ge 1$.

Thus, we found an explicit expression for the partial partition functions of the three-coloring statistical model in the trigonometric limit. The challenging problem is to find the corresponding expression for the general elliptic case.

\subsubsection*{Acknowledgments}

This work was supported in part by the RFBR grant \#07--01--00234. We wish to acknowledge the warm hospitality of the Erwin Schr\"o\-dinger International Institute for Mathematical Physics where the main part of this work was carried out. The first author was also supported in part by the joint DFG--RFBR grant \#08--01--91953. He expresses his gratitude to Profs. H. Boos, F. G\"ohmann and A. Kl\"umper for hospitality at the University of Wuppertal and interesting discussions.

\end{document}